\documentclass[twocolumn]{aastex631}

\usepackage{floatrow}
\usepackage{graphicx}
\usepackage{amsmath,amssymb,amsfonts}
\usepackage{amsthm}
\usepackage{mathrsfs}
\usepackage{booktabs}
\usepackage{wrapfig}

\received{January 8, 2025}
\revised{January 23, 2025}
\accepted{January 31, 2025}

\def\deg{\ifmmode{^\circ}\else{$^\circ$}\fi}
\def\co2{\ifmmode{{\rm CO}_2}\else{CO$_2$}\fi}
\def\h2o{\ifmmode{{\rm H}_2{\rm O}}\else{H$_2$O}\fi}

\graphicspath{{/}}

\begin{document}

\title{The Influence of Stellar Chromospheres and Coronae on Exoplanet Transmission Spectroscopy}

\correspondingauthor{Volker Perdelwitz}
\email{volker.perdelwitz@weizmann.ac.il}

\author[0000-0002-6859-0882]{Volker Perdelwitz}
\affiliation{Department of Earth \& Planetary Sciences \\ Weizmann Institute of Science \\ Rehovot 76100, Israel}

\author[0009-0006-6440-4209]{Adam Chaikin-Lifshitz}
\affiliation{Department of Earth \& Planetary Sciences \\ Weizmann Institute of Science \\ Rehovot 76100, Israel}

\author[0000-0002-9152-5042]{Aviv Ofir}
\affiliation{Department of Earth \& Planetary Sciences \\ Weizmann Institute of Science \\ Rehovot 76100, Israel}

\author[0000-0001-9930-2495]{Oded Aharonson}
\affiliation{Department of Earth \& Planetary Sciences \\ Weizmann Institute of Science \\ Rehovot 76100, Israel}
\affiliation{Planetary Science Institute \\ Tucson 85719-2395, AZ, USA}

\begin{abstract}
\noindent  A main source of bias in transmission spectroscopy of exoplanet atmospheres is magnetic activity of the host star in the form of stellar spots, faculae or flares. However, the fact that main-sequence stars have a chromosphere and a corona, and that these optically thin layers are dominated by line emission may alter the global interpretation of the planetary spectrum, has largely been neglected. Using a JWST NIRISS/SOSS data set of hot Jupiter HAT-P-18~b, we show that even at near-IR and IR wavelengths, the presence of these layers  leads to significant changes in the transmission spectrum of the planetary atmosphere. Accounting for these stellar outer layers thus improves the atmospheric fit of HAT-P-18~b, and increases its best-fit atmospheric temperature from  $536^{+189}_{-101}$~K to  $736^{+376}_{-188}$~K, a value much closer to the predicted equilibrium temperature  of $\sim$852~K. Our analysis also decreases the best-fit abundance of CO$_2$  by almost an order of magnitude. The approach provides a new window to the properties of chromospheres/corona in stars other than our Sun.
\end{abstract}

\keywords{Hot Jupiters, Exoplanet atmospheres, Stellar atmospheres }

\section{Introduction} \label{sec:intro}

\begin{figure}
\includegraphics[width=0.95\textwidth]{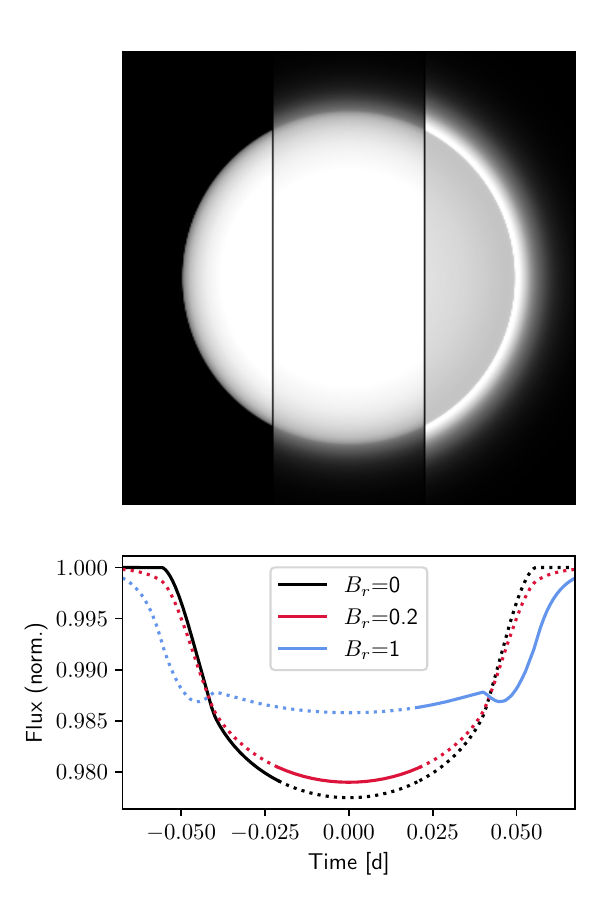}
\caption{Depiction of the effect of chromospheres/coronae on transit shapes. We computed three scenarios of a stellar disk with a photosphere and an optically thin outer layer, with increasing brightness ratio $B_r$ (i.e. the ratio between the brightness of the chromosphere/corona and that of the photosphere) from zero to 0.2 and unity. The lower panel shows the light curve of a planet transiting a star with the three brightness profiles displayed above, all assuming the planetary parameters of HAT-P-18 b. The solid section of each curve highlights the corresponding profile in the panel above.}
\label{figlec}
\end{figure}

\noindent The exceptional capabilities of JWST, along with the development of sophisticated exoplanet atmosphere models, has led to notable advances in the field of atmospheric studies via transmission spectroscopy  \citep{2023Natur.614..653A,2023Natur.614..664A,2023ApJ...943L..10C,2023Natur.614..670F,2023Natur.614..659R, 2024Natur.626..979P, 2024ApJ...962L..20R,2024ApJ...963L...5X}. On the other hand, while JWST data acquired by its various instruments has enabled the discovery of previously undetected molecular species and atmospheric properties, it also made common knowledge that the dominant limiting source of noise for many systems is magnetic activity of the stellar host  in the form of photospheric heterogeneities via the transit light source effect \citep{2023ApJ...959...64H,2023ApJ...955L..22L,2023ApJ...959L...9M,2023ApJ...948L..11M,2024ApJ...970L...2C,2024MNRAS.528.3354F,2024arXiv240919333R,2024arXiv240415505S}.
The outer layers of the stellar atmosphere, the chromosphere and corona, are usually associated with UV and X-ray emission, despite the fact that both have originally been discovered in visible light \citep{c7f422a0-f90a-3de4-a1f5-51f463162fbf,1868RSPS...17...91L}. Including optically thin extended layers like these in the transit model will result in a shallower but wider transit, which cannot be accounted for by standard photospheric transit models without introducing wavelength-dependent changes to the orbital parameters. Fig.\ref{figlec} shows a phenomenological overview of the effect of such an optically thin outer layer on transit shape, with three numerically simulated brightness profiles and their resulting transits, assuming orbital parameters of HAT-P-18~b.
In order to model the occultation of light from all layers of the stellar atmosphere, we developed  Transits Across CHromosphEricaeLy activE Stars (TACHELES) , which combines a photospheric \citep[i.e., ][]{2002ApJ...580L.171M} component with one that describes the extended, optically thin nature of the chromosphere and corona. Since both chromosphere and corona cool via line emission, their resulting spectra and all effects on transit light curves are highly wavelength-dependent \citep{alma992910777903596}.

In this work, we describe our first application of the model to the hot Jupiter HAT-P-18~b.
We start with outlining the TACHELES model in \S\ref{sec:model}, followed by a description of the JWST HAT-P-18~b data reduction and light curve fitting (\S\ref{lcf}), the comparison of multiple atmospheric retrievals (\S\ref{psf}), and conclusions.

\section{The TACHELES model} \label{sec:model}

TACHELES combines a traditional photospheric occultation model with an additional contribution from an exponentially decaying, optically thin, emitting spherical shell - the effective chromospheric/coronal layer. The latter is similar to the X-ray transit model recently described in \cite{2024arXiv241001559K}, but was independently developed here. The photospheric component, $F_p$, is a standard  \cite{2002ApJ...580L.171M} model with a typical quadratic limb darkening law. This component was computed using {\sc batman} \citep{2015PASP..127.1161K}. The second component, $F_c$, is detailed below and is the main contribution of this work.
The total transit light curve from TACHELES as a function of the star-planet separation distance, $d$, is then 
\begin{multline}
 F(r_p,u_1,u_2,\alpha,B_r,H;d)=
 \\ \frac{\alpha}{1+B_r}[{F_p(r_p,u_1,u_2;d)+B_r F_c(r_p,H;d)}],
    \label{eq:Total_LC}   
\end{multline} 

where the four traditional transit parameters are the effective planetary radius, $r_p$, quadratic limb darkening parameters, $u_1$ and $u_2$, and flux normalization, $\alpha$.   The two new chromospheric/coronal parameters are its brightness ratio relative to the photosphere, $B_r$, and scale height, $H$. 

\begin{figure*}[ht!]
\centering\small
\includegraphics[width=0.45\textwidth]{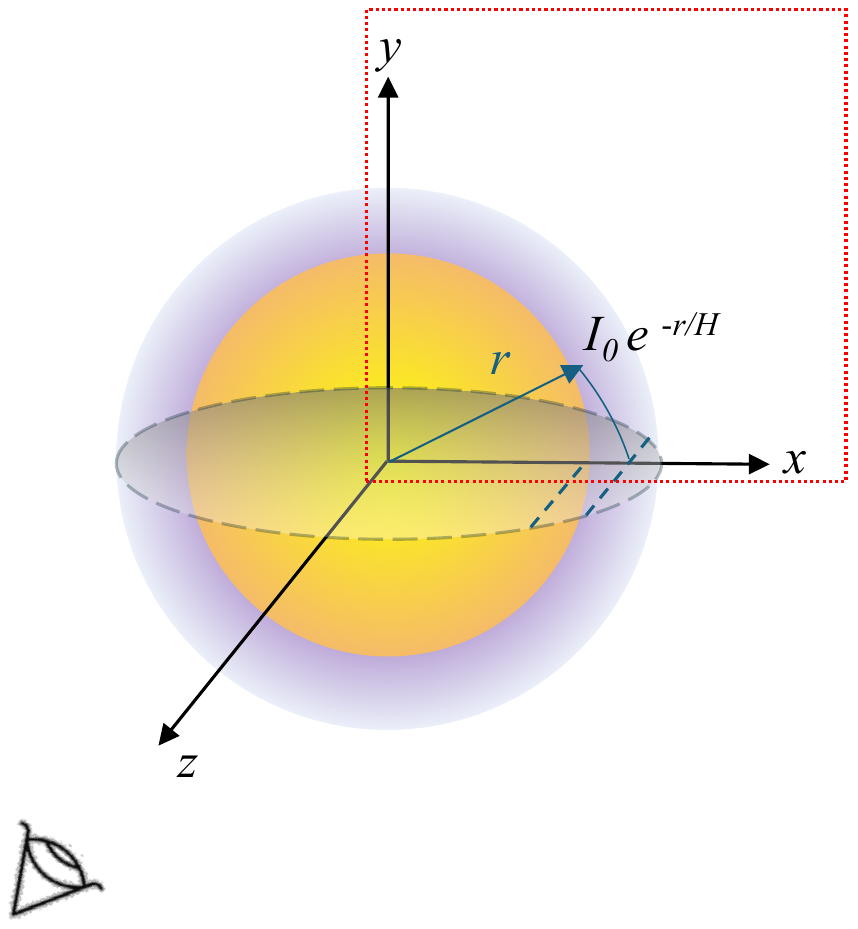} 
\includegraphics[width=0.45\textwidth]{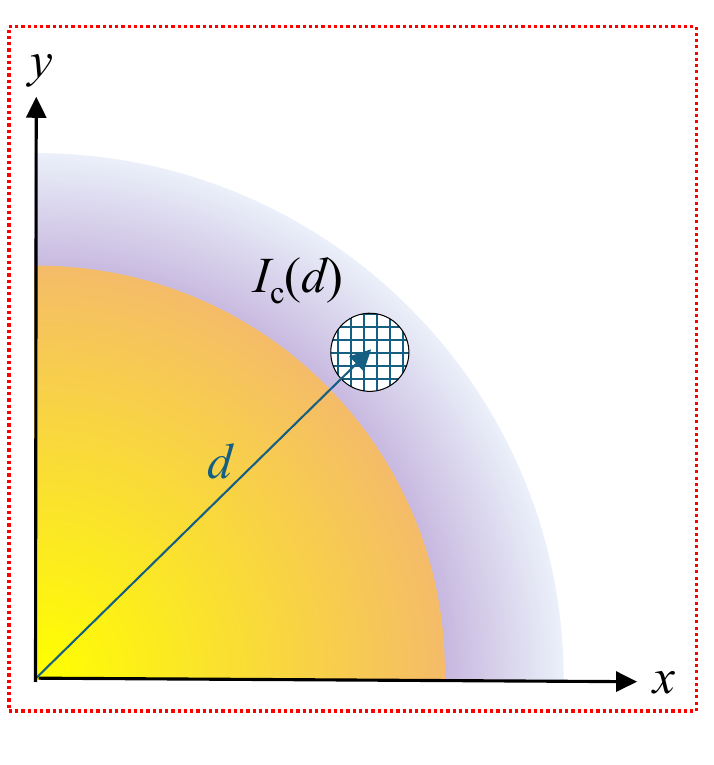}
\caption{The geometry of the observation. Left: In perspective view, the chromospheric/coronal intensity falls off exponentially with radial distance $r$ from the surface of the star. We integrate this quantity along chords parallel to the line of sight (for example, inside and outside the photospheric limb as dashed blue lines). Right: In the $x-y$ plane perpendicular to the observer's line-of-sight, the chromosphere is discretized, and the symmetry about the $z$ axis is used for efficient computation of the apparent intensity on the sky plane $I_c(d)$, where $d$ is the projected distance between the planet and star centers in this plane.\vspace{0.3cm}}
\label{f:diagram}
\end{figure*}

The chromospheric/coronal model below computes the observed occultation profile, starting from the assumed emission function, which varies with distance from the center of the star $r = \sqrt{x^{2}+y^{2}+z^{2}}$,
where $x$ and $y$ form the sky-plane and $z$ is parallel to the line of sight (Fig.~\ref{f:diagram}), all measured in units of the stellar photospheric radius, $r_\star$. 
Consistent with observations \citep{1980ARA&A..18..439L}, 
we describe chromospheric emission with the exponentially decaying function
$I_{0}e^{-r/H}$,
where $H$ is the scale height of the emitting region, and $I_{0}$ is a normalization constant. In Appendix \ref{SDO_TACHELES} we describe observations of the Sun validating that an exponential decay of the outer stellar layer is indeed a good approximation of the observed line-of-sight integrated emission from these optically-thin Solar-atmospheric layers.
To obtain the radial intensity profile in the sky plane as seen by the observer, we numerically integrate this function along the line of sight in the regions not blocked by the photosphere:
\begin{equation}\label{integrating_intensity}
    I_{c}(d, H)=
    \begin{cases}
    \int_{z_0}^{\infty} I_{0}e^{-r/H} \,dz\ , & \hspace{.5cm} \text{if } x\le 1,\\
    \\
    \int_{-\infty}^{\infty}{I_{0}e^{-r/H}} \,dz\ , & \hspace{.5cm} \text{if } x>1.
\end{cases}
\end{equation}
Here $I_c$ is the on-sky chromospheric/coronal brightness, and $d=\sqrt{x^2+y^2}$ is the radial distance on the sky plane between the centers of the star and the planet. Due to  azimuthal symmetry, we can choose $y=0$ without loss of generality, and then $z_0=\sqrt{1-x^{2}}$.  Note, the model includes an inherent discontinuity at the stellar photospheric limb where the column length into the chromosphere is halved (Fig.~\ref{f:diagram}), leading to two cases in Eq.~\ref{integrating_intensity}. We normalize this function such that the integral of $I_c$ over the sky plane is unity.

For our calculations, we perform this integration numerically over a domain of 8000 pixels in the $x$ direction with a stellar radius of 1000 pixels. At this resolution, both the planet and chromosphere (which may have smaller scale height than the planet) are well sampled.

The left panel of Fig.~\ref{f:diagram} demonstrates the geometry of the chromosphere and the integration process.
The spherically symmetric illumination function that depends on radial distance $d$ is reduced to a projected brightness profile $I_c(r)$, where $r$ is the on-sky distance from the center of the star. Rotational symmetry about the $z$-axis allows performing the chord integral (along the dashed lines) in the $x-z$ plane (delineated by the dashed circle).  This allows assigning a brightness value for each point on the sky plane without additional calculations.
We normalize the chromospheric/coronal brightness profile and integrate it over the discretized sky-projected disk of the planet (see Fig. \ref{f:diagram}b) to obtain $F_c$. In Appendix \ref{RetrievalTesting} we describe retrieval tests with simulated data, outlining the expected detection limits of TACHELES as a function of various parameters.

\section{Light curve fitting}
\label{lcf}
We use TACHELES to reanalyze the JWST NIRISS/SOSS  \citep{2023PASP..135i8001D,2023PASP..135g5001A} observations of the hot Jupiter HAT-P-18~b \citep{2022ApJ...940L..35F,2023MNRAS.524..377H,2024MNRAS.528.3354F}, chosen based on wavelength coverage, signal-to-noise, availability, and the level of magnetic activity of the host star. HAT-P-18~b is orbiting a K dwarf of moderate magnetic activity \citep[$\log R'_{\mathrm{HK}}$=-4.86,][]{2024A&A...682A.136C}, and the transmission spectrum of the planetary atmosphere has been shown to be influenced by stellar spots \citep{2024MNRAS.528.3354F}. 
In order to avoid any influence of a difference in reduction of the raw NIRISS/SOSS data, we use the spectrophotometric dataset reduced by \citep[][priv. comm.]{2024MNRAS.528.3354F}. We follow their analysis, deriving white light curves for both orders by coadding the single-bin light curves for each order, i.e. $\lambda>0.85\, \mu\mathrm{m}$ for order 1 and $\lambda<0.85\, \mu\mathrm{m}$ for order 2. The two white light curves are then fit with a \cite{2002ApJ...580L.171M} model using {\sc BATMAN} \citep{2015PASP..127.1161K} and {\sc ULTRANEST} \citep{2021JOSS....6.3001B}, fixing the planetary period to 5.508023~d \citep{2011ApJ...726...52H} and leaving uninformed, wide-open priors for the planetary radius, the two quadratic limb darkening coefficients, the transit midpoint, the inclination, the semi-major axis, a constant baseline offset and a jitter term replacing the error produced by the pipeline, as was done in \cite{2024MNRAS.528.3354F}. In both the white light curve fitting and the subsequent analysis of the single-bin light curves the spot crossing event close to the center of the transit was masked out by removing all data within $0.011$~d of $0.005$~d after $t_0$.
The white light curve fit with TACHELES using the same priors as the individual wavelength bin fits (listed in tab.~\ref{tab1}) constrains neither the brightness ratio nor the scale height. This is to be expected, as the emission of chromosphere and corona is dominated by line emission in only a small subset of wavelength bins, and even among this subset they exhibit scale heights varying by several orders of magnitude. For the same reason, the log-likelihood is improved by the TACHELES fit in white light, although no additional constraints are placed on the scale height and brightness ratio. As described in the following data analysis, these wavelength-dependent parameters vary by orders of magnitude, and cannot be constrained to a single value for the white-light fit.

As the order 1 spectrum has the higher signal-to-noise ratio, we fix the following parameters to the best-fit results from this white light fit to the following values: the transit mid-point $t_0=59743.35341588 d$, the inclination $i=88.62440227^{\circ}$ and the semi-major axis $a=16.44403185$~$r_{\star}$. All results of the white light fit are shown in Tab.~\ref{tabwl} in the Appendix.

The light curve for each individual wavelength bin was fitted with the usual photospheric Mandel-Agol (MA) model using {\sc BATMAN} \citep{2015PASP..127.1161K}, as well as with TACHELES - the sum of a photospheric model and our new coronal/chromospheric model, the total normalized to unity (Eq.~\ref{eq:Total_LC}, priors given in Table~\ref{tab1}). Our fitted parameter choices follow \cite{2024MNRAS.528.3354F}.
For both the photospheric and the TACHELES model, the remaining free parameters are the planetary radius $r_p$, the two limb darkening parameters for a quadratic model $u_1$ and $u_2$, and a normalization coefficient $\alpha$. The TACHELES model has two additional parameters, the brightness ratio between chromosphere/corona and photosphere $B_r$, and the scale height $H$. 
We normalize the distances $r_p$, $H$, and $d$ by the stellar radius, $r_\star$.

The fit was performed with {\sc ULTRANEST} \citep{2021JOSS....6.3001B} using uniform priors for $r_p$, $\alpha$, $\log B_r$ and $\log H$. For $u_1$ and $u_2$ we used gaussian priors with a width of 0.2 \citep{Patel_2022} centered on the values predicted with {\sc exotik-ld} based on {\sc PHOENIX models} \citep{david_grant_2022_7437681,2013A&A...553A...6H}.
All prior ranges are displayed in Table~\ref{tab1}. 
In Appendix \ref{TACHELES_corner} we show the correlation among all the fitted parameters for a wavelength with well-constrained $H, B_r$.

\begin{figure*}
\includegraphics[width=0.75\textwidth]{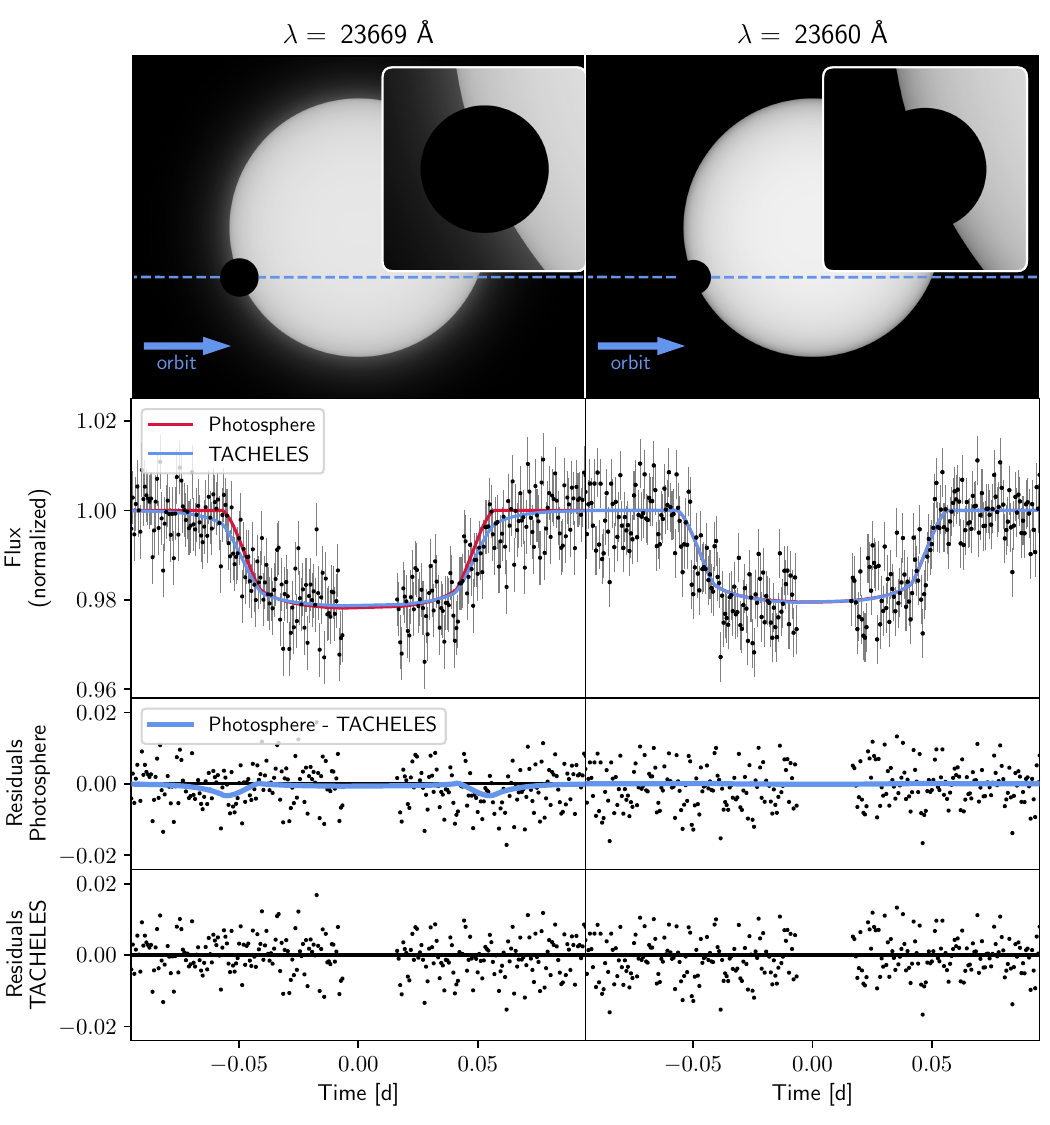}
\caption{Example of a wavelength bin exhibiting significant chromospheric/coronal emission (left column) and an adjacent nearly unaffected wavelength bin (right column). Top row:Illustration of the best-fit TACHELES models. Second row:
The observed light curve along with the resulting model fits. Third row: Residuals from the traditional Mandel-Agol photospheric model fit, highlighting the difference between MA and TACHELES models. Bottom row: Residuals from the TACHELES model.\vspace{0.3cm}}
\label{fig1}
\end{figure*}

Fig.~\ref{fig1} shows the TACHELES model for a wavelength bin ($\lambda=23669 \text{\r{A}}$) with a significant contribution of the outer stellar layers compared to an adjacent wavelength bin ($\lambda=23660 \text{\r{A}}$) that exhibits a light curve consistent with purely photospheric emission, along with the observed light curves and model fits.

\begin{figure}[t]
\centering
\includegraphics[width=\textwidth]{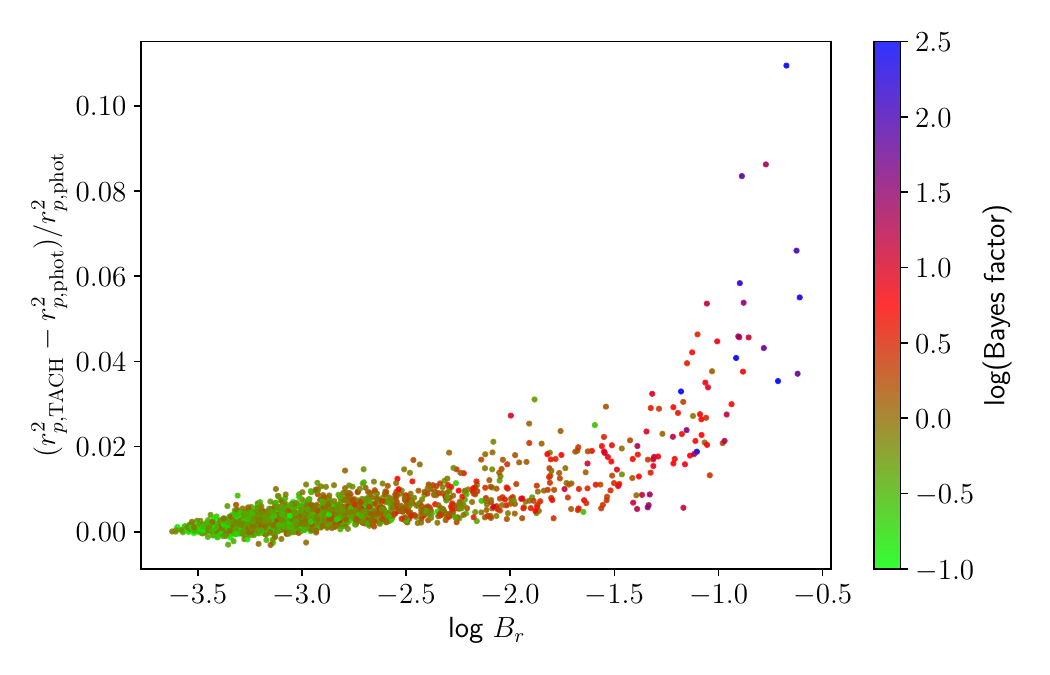}
\caption{Relative difference in transit depth derived for the TACHELES model radius ($r_{p,\mathrm{TACH}}$) and the photospheric model radius ($r_{p,\mathrm{phot}}$) as a function of the TACHELES brightness ratio. The color coding denotes the Bayes factor.}\label{fig:deltaR}
\end{figure}

We found that, according to the Jeffrey's scale \citep{Jeffreys1939-JEFTOP-5,WASSERMAN200092}, 24 bins show substantial evidence preferring TACHELES over the photospheric fit (Bayes factor $\geq$3), and 6 bins show strong evidence (Bayes factor $\geq$10).
Fig.~\ref{fig:deltaR} shows the change in planetary radius with regard to the TACHELES brightness ratio, color-coded by the Bayes factor. Wavelength bins with strong evidence for TACHELES exhibit a larger deviation in radius, as well as a higher brigthness ratio. We also note that we do detect prominent lines associated with stellar chromospheres and coronae, such as the Fe X line, which was one of the spectral signatures that led to the discovery of the solar corona \citep{1939NW.....27..214G,2014FrASS...1....2P}.

\renewcommand{\arraystretch}{1.1}
\begin{table}[ht]
\caption{TACHELES and photospheric light curve fit priors for {\sc ULTRANEST}. $r^\mathrm{WL}_p$ denotes the radius derived from the white light fit.}\label{tab1}
\begin{tabular}{@{}lcc@{}}
\toprule
Parameter&Priors (Photosphere)& Priors (TACHELES)\\
\midrule
$r_p$&$\mathcal{U}$\{0.85 $r^{WL}_p$,1.15  $r^{\mathrm{WL}}_p$\}&$\mathcal{U}$\{0.85 $r^\mathrm{WL}_p$,1.15  $r^{WL}_p$\}\\
$\alpha$&$\mathcal{U}$\{0,10000\}&$\mathcal{U}$\{0,10000\}\\
$u_1$&$\mathcal{N}\{u_1^{exotik-ld},0.2\}$&$\mathcal{N}\{u_1^{exotik-ld},0.2\}$\\
$u_2$&$\mathcal{N}\{u_2^{exotik-ld},0.2\}$&$\mathcal{N}\{u_2^{exotik-ld},0.2\}$\\
$\log B_r$&-&$\mathcal{U}$\{-6,1\}\\
 $\log H$&-&$\mathcal{U}$\{-3,0\}\\
\botrule
\end{tabular}
\end{table}
\renewcommand{\arraystretch}{1}

These bins are distributed across the entire wavelength range, with scale heights ranging from $0.07$ to $0.31$ (in stellar radius units), indicating that indeed both chromosphere and corona influence the transit shape at distinct wavelengths. The brightness ratios of these bins relative to the photosphere have a range of $0.04$ to $0.25$. Both of these ranges are in agreement with the expected values for main-sequence stars \citep{1980ARA&A..18..439L, alma992910777903596}.
Since the presence of emitting material near the limb of the star will dilute the apparent photospheric transit depth in the same manner as an unocculted active region/third light would \citep{2018ApJ...853..122R}, the relative difference in radii between TACHELES and the purely photospheric model is positive for $98$\% of bins, and there is a clear wavelength-dependency, as shown in the lower panel of fig.~\ref{fig:pos}. In wavelength bins with a significant contribution of the chromosphere or corona, however, the photospheric model can underestimate the planetary radius by as much as $5.3$\%. An example corner plot is displayed in fig.~\ref{fig:corner} in the Appendix.

\section{POSEIDON retrieval}
\label{psf}
In order to test the influence of this deviation on the derived properties of the planetary atmosphere of HAT-P-18~b, we fit the photospheric and TACHELES transmission spectra with the {\sc POSEIDON} retrieval code \citep{2017MNRAS.469.1979M,MacDonald2023} assuming three stellar configurations (no spots, single spot, two heterogeneities), and a planetary configuration for hot jupiters, as described in \cite{2024MNRAS.528.3354F}, i.e. a pressure-temperature parameterization from \cite{2009ApJ...707...24M} and a cloud model based on \cite{2017MNRAS.469.1979M} with the same setup as \cite{2024MNRAS.528.3354F}. All priors are listed in Tab.~\ref{tab:2}.

\renewcommand{\arraystretch}{1.}

\begin{table}
\footnotesize
\caption{Priors for the {\sc POSEIDON} fit. The parameters are the same as in \citep{2024MNRAS.528.3354F}. $r^{WL}_p$ denotes the radius derived from the white light fit.}\label{tab:2}
\begin{tabular}{@{}lc@{}}
\toprule
Parameter&Priors \\
\midrule
\bf{Composition} & \\
log $X_i$ & $\mathcal{U}$\{-12,-1\}\\
 & \\
\bf{P-T profile} & \\
$\alpha_{1,2}$ & $\mathcal{U}$\{0.02,2\}\\
$\log P_{1,2}$ & $\mathcal{U}$\{-8,2\}\\
$\log P_{3}$ & $\mathcal{U}$\{-2,2\}\\
$T_\text{ref}$ & $\mathcal{U}$\{300,2000\}\\
&\\
\bf{Aerosols}&\\
$\log a$ & $\mathcal{U}$\{-4,8\}\\
$\gamma$ & $\mathcal{U}$\{-20,2\}\\
$\log P_\text{cloud}$ & $\mathcal{U}$\{-8,2\}\\
&\\
\bf{Stellar (one het.)}&\\
$f_\text{het}$&$\mathcal{U}$\{0,0.5\}\\
$T_\text{het}$&$\mathcal{U}$\{3500,1.2 $T_\star$\}\\
$T_\text{phot}$&$\mathcal{N}$\{$T_\star$,$\sigma_{T_\star}$\}\\
&\\
\bf{Stellar (two het.)}&\\
$f_\text{fac}$&$\mathcal{U}$\{0,0.5\}\\
$f_\text{spot}$&$\mathcal{U}$\{0,0.5\}\\
$T_\text{fac}$&$\mathcal{U}$\{$T_\star-3\,\sigma_{T_{\star}}$,1.2 $T_{\star}$\}\\
$T_\text{spot}$&$\mathcal{U}$\{3500,$T_\star+3\,\sigma_{T_{\star}}$\}\\
$T_\text{phot}$&$\mathcal{N}$\{$T_\star$,$\sigma_{T_\star}$\}\\
&\\
\bf{Other} & \\
$r_{p,\text{ref}}$&$\mathcal{U}$\{0.85 $r^{WL}_p$,1.15 $r^{WL}_p$\}\\
\botrule
\end{tabular}
\end{table}
\renewcommand{\arraystretch}{1}

\begin{figure*}[ht!]
\centering
\includegraphics[width=0.85\textwidth]{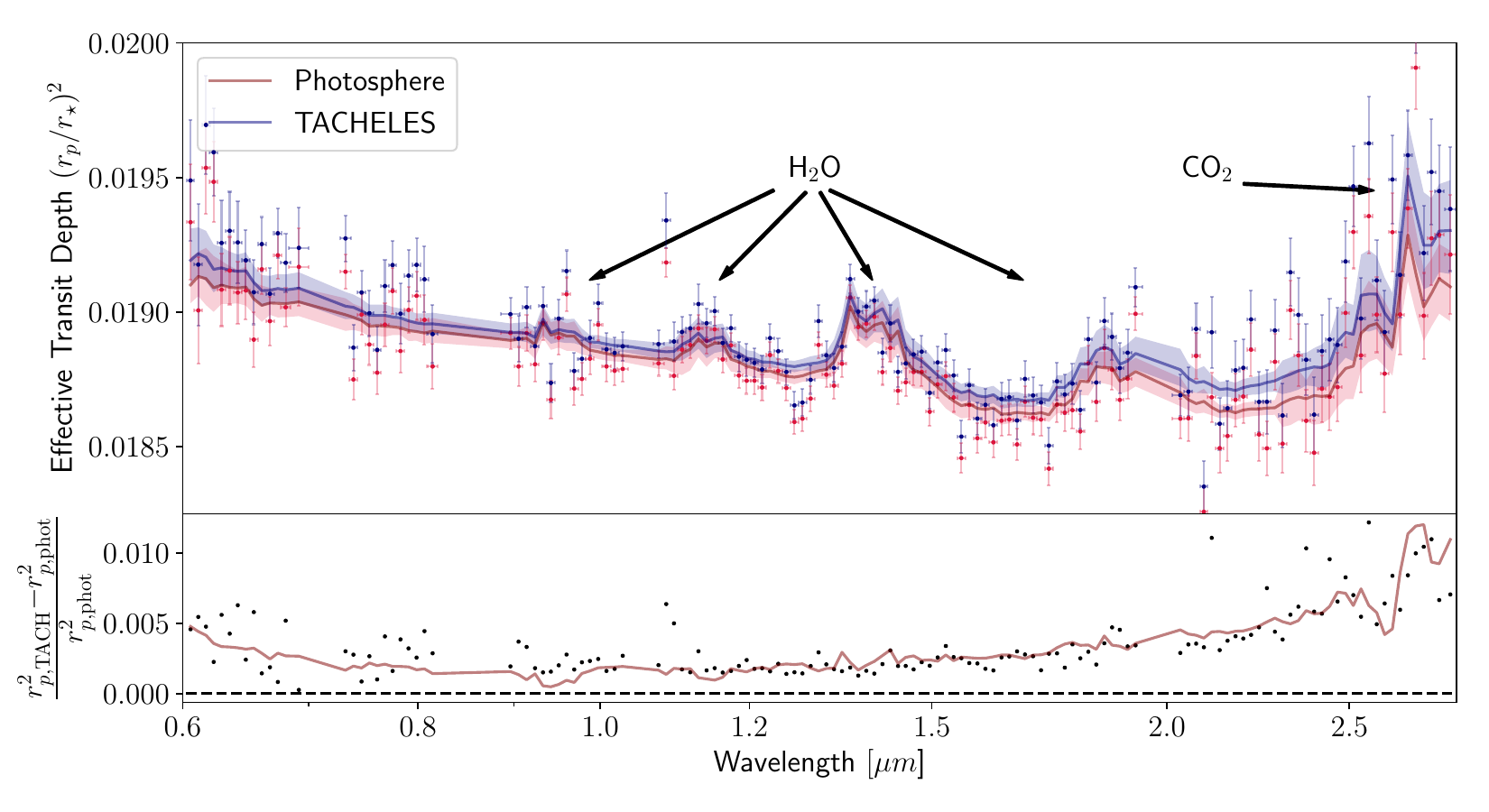}
\caption{Transmission spectra of HAT-P-18~b and {\sc POSEIDON} fits, binned to a resolution of 100. The POSEIDON models (one active region) are plotted as solid lines and $1\sigma$ errors are marked by the shaded regions. The lower panel shows the relative point-by-point difference between the transit depths of TACHELES and the photosphere (black dots) as well as that of the two POSEIDON fits (solid line).}\label{fig:pos}
\end{figure*}

Since \cite{2024MNRAS.528.3354F} found no evidence for any other species, we ran the fit for each spot configuration including only Na, H$_2$O, CO$_2$, with both retrieved spectra binned to a resolution of 100. For both the TACHELES-based transmission spectrum (Fig.~\ref{fig:pos}) and the photospheric one we find that the POSEIDON model with a single heterogeneity yields the highest log likelihood. 
In the two heterogeneity case, we found that allowing a broad prior on the area coverage of faculae reduces the quality of fit, so the data prefers a small second heterogeneity. 
Since {\sc POSEIDON} does not yet include the possibility of logarithmic priors for this parameter, we confirmed that the addition of faculae does not improve the fit by progressively decreasing the prior range for this parameter effectively forcing it to a value close to zero. Indeed, in that process, the likelihood increases steadily and converges to the values of the one-heterogeneity model at the prior range of [0,0.01]. We conclude that this illustrates that faculae do not contribute to the fit in both the photospheric- and TACHELES-derived spectra. Table~\ref{tab2} shows the resulting best-fit values with $1\sigma$ uncertainties.

\renewcommand{\arraystretch}{1.3}
\begin{table*}[ht]
\centering
\caption{Atmospheric fits of HAT-P-18~b. The full parameter set for each fit is available in at \url{https://zenodo.org/records/13361162}, along with the corner plots. The best fitting models for both the photospheric and TACHELES data are those with a single heterogeneity, marked in bold.}\label{tab2}
\begin{tabular}{lccccc}
\toprule
Configuration&$T_{\text{ref}}$ [K] & log~H$_2$O&log~CO$_2$&log~Na&ln\,Z\\
\toprule
Photosphere&$395.8^{+69.1}_{-49.3}$&$-2.97^{+0.51}_{-0.48}$&$-4.59^{+0.98}_{-2.88}$&$-6.95^{+2.21}_{-3.09}$&$984.36\pm0.32$\\
{\bf Photosphere (one het.)}&$536.0^{+188.9}_{-100.8}$&$-3.89^{+0.47}_{-0.46}$&$-4.75^{+0.66}_{-0.74}$&$-9.07^{+1.71}_{-1.85}$&$984.72\pm0.31$\\
Photosphere (two het.)&$516.7^{+105.4}_{-86.5}$&$-3.90^{+0.47}_{-0.41}$&$-4.73^{+0.65}_{-0.70}$&$-9.10^{+1.58}_{-1.78}$&$984.00\pm0.33$\\
TACHELES&$471.9^{+96.7}_{-78.7}$&$-3.60^{+0.51}_{-0.48}$&$-4.35^{+0.67}_{-0.77}$&$-5.20^{+0.86}_{-1.41}$&$985.05\pm0.34$\\
{\bf TACHELES (one het.)}&$720.8^{+376.4}_{-188.0}$&$-4.75^{+0.66}_{-0.69}$&$-5.42^{+0.90}_{-0.92}$&$-8.28^{+1.34}_{-2.11}$&$989.46\pm0.31$\\
TACHELES (two het.)&$679.8^{+230.3}_{-153.0}$&$-4.59^{+0.54}_{-0.49}$&$-5.17^{+0.78}_{-0.67}$&$-8.00^{+1.22}_{-1.84}$&$987.45\pm0.33$\\
\botrule
\end{tabular}
\end{table*}

\renewcommand{\arraystretch}{1}

The main difference is the increase in the model atmospheric temperature from $536.0^{+188.9}_{-100.8}$~K for the photospheric fit to $720.8^{+376.4}_{-188.0}$~K for TACHELES, a value which is close to the equilibrium temperature for HAT-P-18~b  of $852$~K. The increase in the retrieved temperature may be understood as a consequence of adding the chromospheric/coronal light source, which requires a larger effective planetary radius to match the transit depth. The larger radius, in turn, favors a puffier atmosphere - having higher-temperature and/or lower mean molecular weight - and both are now favored by TACHELES. This suggests that the disregard of the stellar outer layers could explain the surprisingly low atmospheric temperatures reported for the majority of exoplanets, relative to their equilibrium temperatures \citep{2020ApJ...893L..43M}.
Additionally, the TACHELES analysis found tentative evidence for a decrease in abundance for both water ($2\sigma$) and CO$_2$ ($1\sigma$) relative to the photospheric model. 
However, the latter finding is based on a singular feature at the edge of the spectrum, and it has been found that the abundance of CO$_2$ is hard to constrain with NIRISS/SOSS data \citep{2023MNRAS.524..817T} and our findings do not resolve the deviation of H$_2$O from expected abundance.

\section{Plasma model for the chromspheric/coronal emission}
\begin{figure*}[ht!]
\centering
\includegraphics[width=0.8\textwidth]{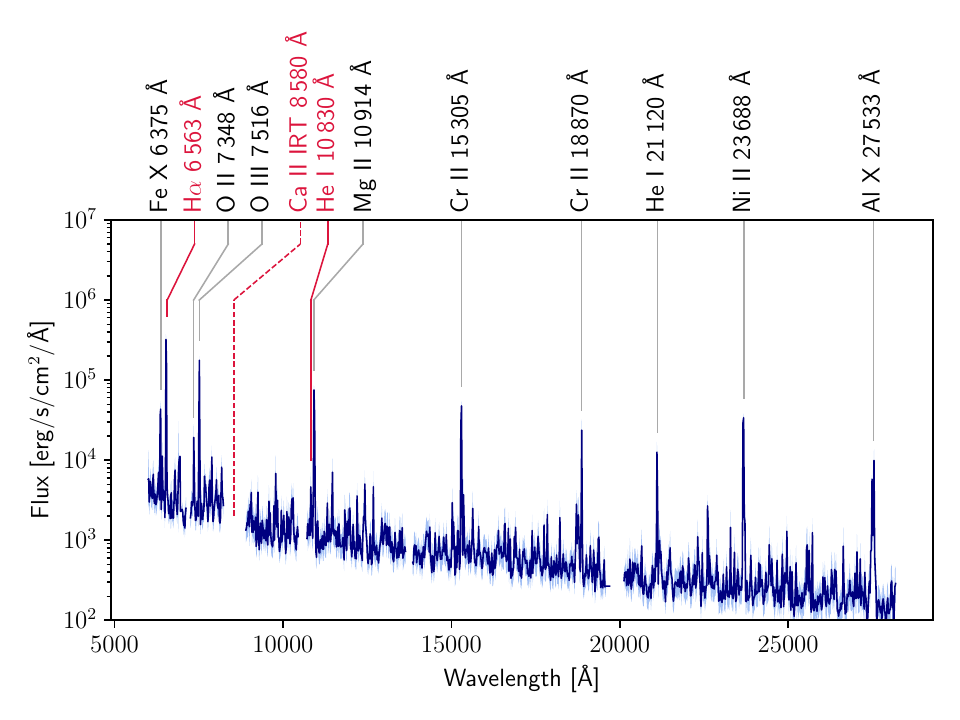}
\caption{Chromospheric/coronal emission spectrum of HAT-P-18 derived by multiplication of the TACHELES model brightness ratios with a PHOENIX photospheric spectrum. The solid blue line shows the spectrum smoothed to the instrumental resolution, along with the $1\sigma$ error region in light blue. Preliminary line identifications are written in black, and marginal/non-detection identifications of notable species are written in red.}\label{f:cspectrum}
\end{figure*}

In addition to removing biases in temperature and composition from the determination of the atmospheric properties of the planet, TACHELES also allows for the study of the outer layers of the stellar atmosphere, i.e., the chromosphere and corona. A full model of the chromospheric and coronal emission of HAT-P-18 should include, besides the plasma emission (E-corona) with individual metallicities and differential emission measures, the F- and K-corona \citep{2009soco.book.....G} - all beyond the scope of this publication. Instead, we present a preliminary identification of the clear emission features in the resulting spectrum of the non-photospheric component of TACHELES (Fig.~\ref{f:cspectrum}), which is derived by multiplication of the TACHELES brightness ratio with a PHOENIX photospheric model \citep{2013A&A...553A...6H}.
Since there are no sufficient UV and X-ray data of HAT-P-18 to compute a differential emission measure, we can only compare our findings to a plasma model based on Solar data. However, this comparison also serves as a sanity check for our method, as we expect to be able to identify all clear emission lines in the spectrum.
In order to carry out a preliminary identification of features in the chromospheric/coronal spectrum, we compare it to a model based on the differential emission measure of a Solar prominence \citep{1997SoPh..175..411W} with a temperature range of $T=10^{4.1}$ to $10^{6.2}$~K, and assuming solar abundances, allowing us to compute a {\sc CHIANTI} \citep{1997A&AS..125..149D,2013ascl.soft08017D,2021ApJ...909...38D} spectrum for the wavelength range of the NIRISS observation. The temperature range and metallicities of the plasma is in agreement with the values derived for other moderately active K dwarfs \citep{2006ApJ...643..444W}.
We derive a line list from the CHIANTI model and identify matches with the TACHELES chromospheric model by, within a NIRISS spectral resolution element, maximizing $\mathcal{I}/\Delta\lambda$, where $\mathcal{I}$ is the CHIANTI line intensity and $\Delta\lambda$ is its wavelength offset relative to the TACHELES line.  

{\sc CHIANTI} does not take into account temperatures below $10^4$~K, so it is not currently possible to probe the lower range of chromospheric temperatures with this tool.

The TACHELES chromospheric/coronal spectrum exhibits a total of 9 lines with a significance of $\geq5\sigma$, for all of which we can identify counterparts in the line list provided by the {\sc CHIANTI} model. These preliminary line identifications are denoted at the top of Fig.~\ref{f:cspectrum} and include transitions from singly-ionized atoms at lower temperatures to highly-ionized ones, which must originate from the stellar corona.
Some chromospheric features predicted by {\sc CHIANTI} are marginal or non-detections in our analysis, either because they fall in one of the wavelength regions with contamination (e.g., Ca II IRT), or do not rise above our detection threshold. The He~I triplet at $10830$~\AA{} is only marginally detected, and while there is a strong emission peak near the H$\alpha$ line, it is 34~\AA{} bluewards of the line center, which is larger than can be explained by the spectral precision of NIRISS/SOSS at this wavelength. The latter two findings may be related to material that has been found to be evaporating from the atmosphere of HAT-P-18~b \citep{2021ApJ...909L..10P,2022ApJ...940L..35F,2024MNRAS.528.3354F} and will require further investigation.

\section{Conclusions and Outlook}
We find that the presence of a stellar chromosphere and corona can cause a bias in the derivation of exoplanet atmosphere transmission spectra if not accounted for by an appropriate transit model. In the case of HAT-P-18~b, these changes are an increase of the atmospheric temperature from $536$~K to $721$~K, as well as a drop in the mixing ratios of both H$_2$O and CO$_2$ by almost an order of magnitude. Despite HAT-P-18 being only slightly more active than the Sun  \citep[$R'_{\mathrm{HK}}=-4.86$,][]{2024A&A...682A.136C}, the differences in atmospheric parameters derived with TACHELES are significant compared to a purely photospheric transit model. Comparing all past and scheduled JWST transmission spectroscopy targets \citep{2022RNAAS...6..272N} with archival $R'_{\mathrm{HK}}$ values \citep{2024A&A...683A.125P} shows that the majority of host stars are more active than HAT-P-18, suggesting that the influence of the outer stellar atmosphere should be greater in investigations of the planetary atmospheric properties of these targets.
In the future, we will investigate whether  additional information on the chromospheric and coronal emission in the X-ray and UV regime can be used to improve the constraints derived with TACHELES. For high SNR observations, such as from JWST, adding the upper components of the stellar atmosphere to the modeling offers two benefits beyond a better understanding of the planetary atmosphere: firstly, it provides a means for estimating the high-energy irradiation of exoplanets, relevant for habitability \citep{2016ApJ...820...89F}. Secondly, it opens a new window into the coronal and chromospheric layers of stars other than our Sun.
We plan to reinvestigate the transmission spectra of targets with public JWST data with TACHELES, particularly those where it has been found that multiple observations yield disagreeing results \citep[e.g.][]{2023ApJ...959L...9M}. Furthermore, we will develop a more sophisticated approach in which we fit all light curves simultaneously and replace the free parameters of scale height and brightness ratio for the individual light curves with a plasma model differential emission measure and scale height distribution common to all wavelength bins.

\section*{Acknowledgements}
The authors thank the anonymous referee for their detailed and constructive comments. The authors thank Ryan MacDonald for his help with the use of {\sc POSEIDON} and Marylou Fournier-Tondreau for providing us with the set of reduced spectroscopic light curves.
V. Perdelwitz was funded by the Helen Kimmel Center for Planetary Science Prize Postdoctoral Fellowship and the Dean's Fellowship of the Department of Chemistry at the Weizmann Institute of Science.
This study was supported by the Minerva Center for Life Under Extreme Planetary Conditions \#13599 and Minerva Grant \#151010.
This work is based on observations made with the NASA/ESA/CSA James Webb Space Telescope. The data were obtained from the Mikulski Archive for Space Telescopes at the Space Telescope Science Institute, which is operated by the Association of Universities for Research in Astronomy, Inc., under NASA contract NAS 5-03127 for JWST. These observations are associated with the Early Release Observations.
CHIANTI is a collaborative project involving George Mason University, the University of Michigan (USA), the University of Cambridge (UK), and NASA Goddard Space Flight Center (USA).
The SDO data is courtesy of the NASA/SDO and the AIA, HMI, and EVE Science Investigation Teams.

\appendix

\section{TACHELES validation with SDO images} \label{SDO_TACHELES}

\begin{figure}[ht!]
\includegraphics[width=0.6\textwidth]{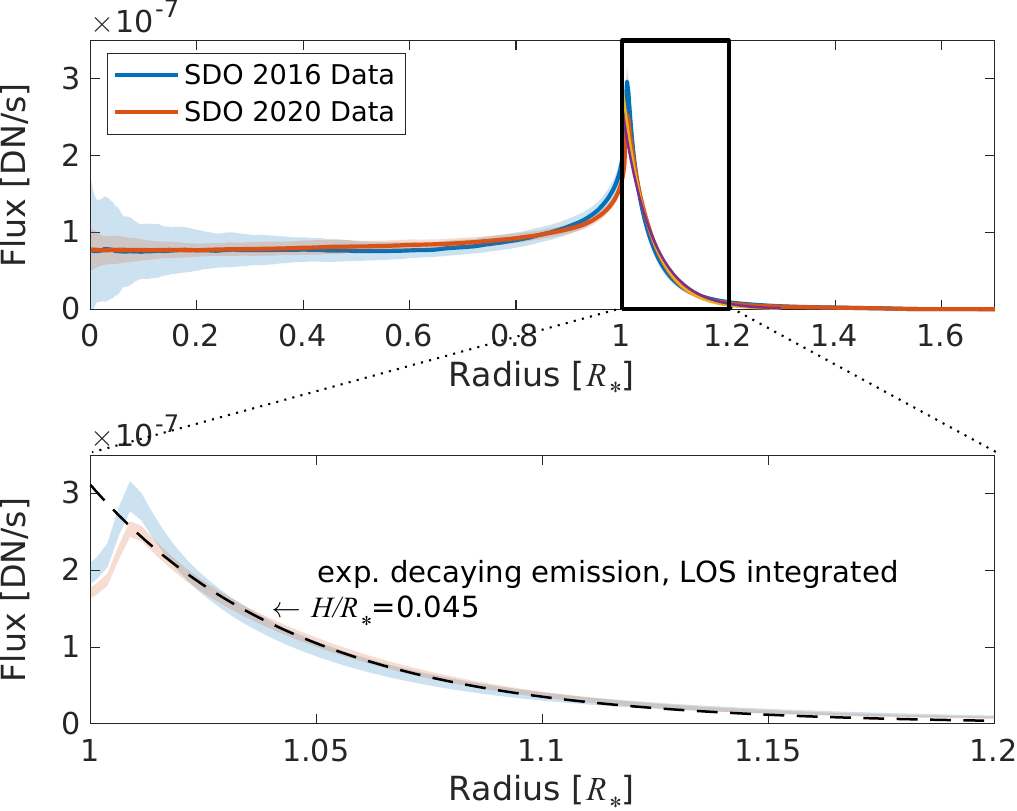}
\caption{Solar Dynamics Observatory \citep{2012SoPh..275....3P} measurements of the Sun's brightness distribution, azimuthally averaged. The data was acquired at 17.1 nm. As highlighted in the enlarged panel, exterior to the solar disk, an exponentially decaying emission law integrated along the line-of-sight of the observer (dashed line) is consistent with the 2020 SDO observations.}
\label{f:sdo}
\end{figure}
To motivate our choice of an exponential distribution of light emission exterior to the photosphere, we analyzed data acquired by the Solar Dynamics Observatory \citep{2012SoPh..275....3P}, averaged separately over two phases of the solar cycle, during 2016 (near solar minimum) and 2020 (near solar maximum).   Figure \ref{f:sdo} shows the brightness distribution, averaged in radial bins.  To correct for the varying distance to the Sun, each observation was normalized in radius according to the stellar photospheric radius as provided in the fits header. For comparison, we also show the profile of an exponentially decaying emission law with a scale-height of 0.045 times the solar radius, which has been integrated along the line-of-sight of the observer.  The agreement with this curve demonstrates that over most of the region exterior to the photosphere, the light distribution at these wavelengths is consistent with an exponential model, as assumed here. We note that in appendix \ref{RetrievalTesting} we show that the size of the planet may set the limit on our ability to discern geometrical details in the chromosphere/corona (e.g., the scale factor $H$) - so a large planet like HAT-P-18\,b may not be sensitive to the small deviation from exponentially decaying shell very close to the photosphere's edge shown in fig. \ref{f:sdo}, but a smaller planet and/or more extended stellar atmospheres may require a more realistic chromospheric/coronal model.

\section{TACHELES detection limits with simulated data} \label{RetrievalTesting}
To test our model and fitting approach, specifically the ability to constrain various system parameters with changing SNR, we generated synthetic light-curves using a set of assumed parameters, varying the brightness ratio, $B_r$, and the scale-height, $H$. The sampling times, orbital period, and other system parameters were chosen to match those of the observations. Random Gaussian noise was added to the light-curves, with varying standard of deviations, corresponding to the inverse per-point signal-to-noise (SNR). We then fit the synthetic light curves using the same approach as was used for the observations.  

\begin{figure}[ht!]
\centering\small
\begin{tabular}{lll}
\bf  a & \bf \ \ \ b & \bf \ \ \ c\\
\hspace*{-6ex}\includegraphics[width=0.35\textwidth]{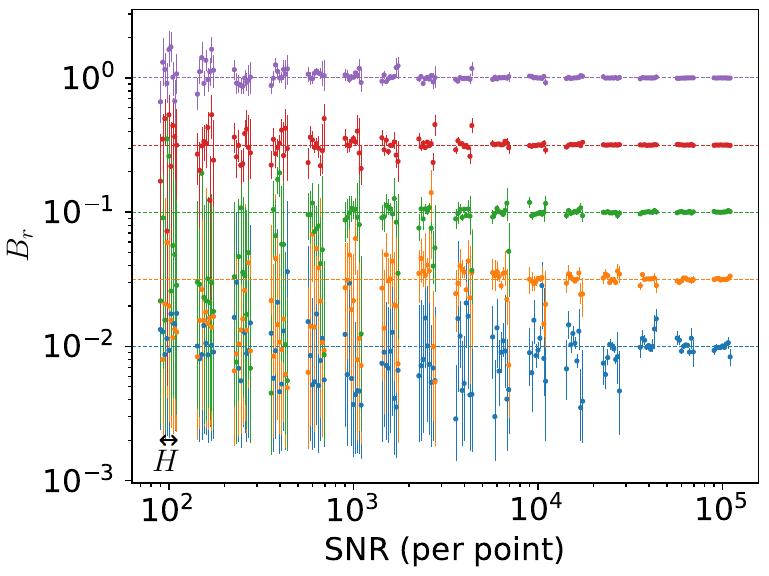} &
\hspace*{-3ex}\includegraphics[width=0.35\textwidth]{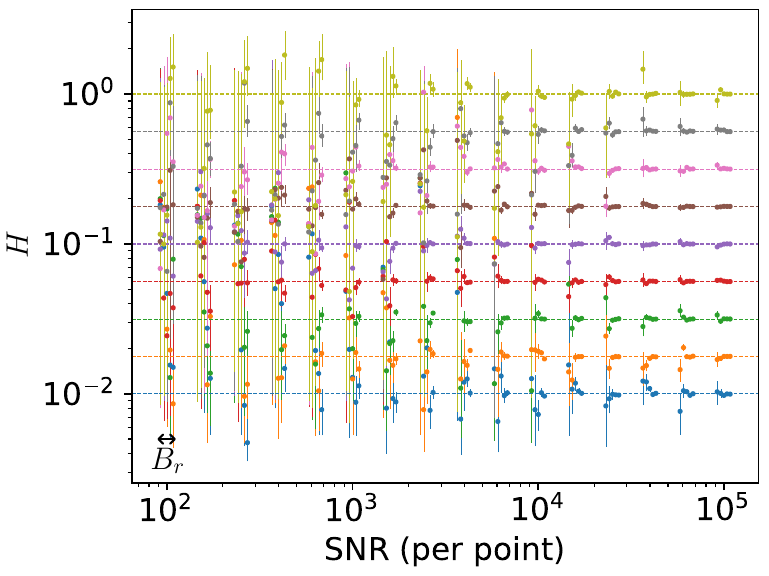}& 
\hspace*{-2ex}\includegraphics[width=0.35\textwidth]{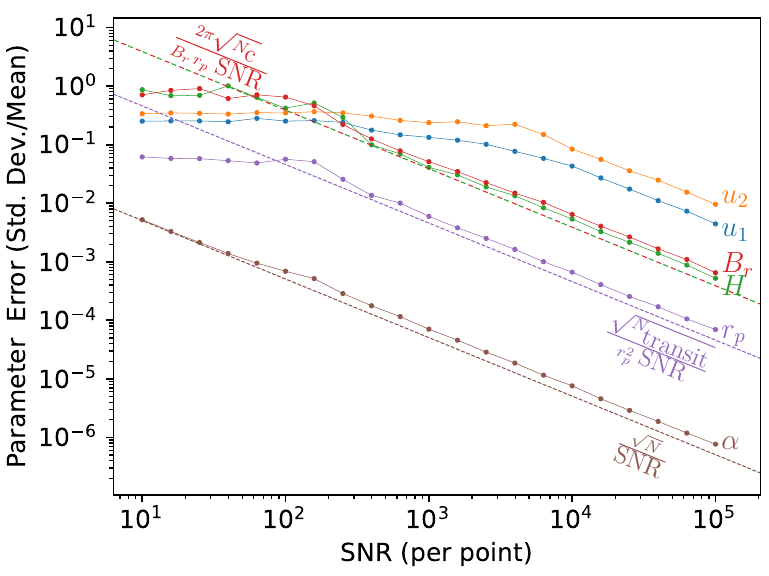}\\

\end{tabular}
\caption{Results of the retrieval tests. \textbf{a} Accuracy of the brightness ratio as a function of SNR. Points with error bars denote the retrieved values and uncertainties for different actual values shown in colors. For each value of $B_r$ and SNR, multiple scale heights were tested, shown with small horizontal offsets. \textbf{b} Same as a, but for the dependence of scale height retrievals on SNR, for different brightness ratios. \textbf{c} 
Scaling of parameter errors with SNR for example choices of $B_r=0.32$, $H=0.1$, and $r_p=0.135$. The predicted convergence lines of $\alpha$, $r_p$, $B_r$, and $H$ are marked as dashed lines, and annotated with simple expressions that predict well the relative errors in the parameters. We ran tests for SNR$=10^{\{2,2.2,2.4,...,5\}}$, $B_r=10^{\{-2,-1.5,-1,-0.5,0\}}$, and $H=10^{\{-2,-1.75,-1.5,...,0\}}$}.
\label{f:tests}
\end{figure}

Figure~\ref{f:tests} shows the results of these tests, plotting the best-fitting values of the derived parameters and their fitting uncertainties. As would be expected, we find that the fitting accuracy depends on the brightness ratio (Fig.~\ref{f:tests}a); at $B_r=1$ useful fits are obtained at an SNR$\sim$100; at $B_r=0.1$ this threshold increases to an SNR$\sim$1000. At the lowest brightness ratio test of $B_r=0.01$, the required SNR is $>10^4$, better than available data from current observations, placing a lower limit on our ability to detect faint non-photospheric contributions in this system.  

A weaker dependence is seen in our ability to retrieve the scale height (Fig.~\ref{f:tests}b). Over the range tested of $H$ between $10^{-2}$ and 1, the intermediate scale-heights near $10^{-1}$ are most easily measured. This is attributed to the size of the occulting planet ($r_p\approx0.019$): large scale heights ($H\gg{}r_p$) are harder to detect since the light emitted by the chromospheric/coronal layer is diffuse: the planet occults only a small fraction of its light. At small scale heights ($H\ll{}r_p$), the chromospheric/coronal emission can be approximated by a thin ring, and the planet occults $\sim B_r r_p/\pi$ of the light, independent of $H$. 
This may be understood as very thin ring adding flux in a smaller number of data points.

The scaling of the errors with SNR (Fig.~\ref{f:tests}c) follow simple statistical estimates, wherein the uncertainties are proportional to $N^{1/2}/\text{SNR}$, where $N$ is the number of points that contribute to the effect \citep{adamsthesis}. $\alpha$ is constrained by all points, so its uncertainty indeed scales simply by $N^{1/2}/\text{SNR}$. For the relative error in the planetary radius there is an additional factor of the area ratio of the planet to the star, $r_p^{2}$, and the number of contributing transit points is $N_\textrm{transit}$. For the errors in the coronal/chromospheric parameters $B_r$ and $H$, the number of points is reduced to $N_c\sim H N_\textrm{transit}$. The amplitude of the effect is reduced, and hence the errors grow by a factor of $B_r$. For $r_p\sim{}H$, the relative area of the overlap of the planet with chromosphere/corona may be roughly estimated as $r_p H/(2\pi H)=r_p/(2\pi)$. As can be seen in the Figure, using example values, these rough estimates provide a good description of the amplitude and scaling of the errors in the model parameters.

\section{Results from the white light curve fits}
Tab.~\ref{tabwl} shows the results of the white light curve fits using both the photospheric model and TACHELES.

\begin{table}
\footnotesize
\caption{Results of the White light curve fit.}\label{tabwl}
\begin{tabular}{@{}lcccc@{}}
Parameter & photosphere O1 & photosphere O2 & TACHELES O1 & TACHELES O2\\
\toprule
\toprule
$r_p$&$0.13761\pm0.00028$&$0.13855^{+0.00053}_{-0.00056}$&$0.13765^{+0.00035}_{-0.00031}$&$0.13847^{+0.00050}_{-0.00053}$\\
$u_1$&$0.285\pm0.029$&$0.566^{+0.037}_{-0.036}$&$0.286^{+0.028}_{-0.030}$&$0.554^{+0.025}_{-0.033}$\\
$u_2$&$0.165^{+0.055}_{-0.053}$&$0.048^{+0.072}_{-0.070}$&$0.172^{+0.056}_{-0.053}$&$0.075^{+0.066}_{-0.049}$\\$t_0$&$59743.35342\pm0.00002$&$59743.35342\pm0.00003$&$59743.35342\pm0.00002$&$59743.35342\pm0.00003$\\
$i$&$88.62\pm0.04$&$88.59^{+0.06}_{-0.05}$&$88.63^{+0.04}_{-0.04}$&$88.60\pm0.05$\\
$a$&$16.50\pm0.06$&$16.44\pm0.08$&$16.51^{+0.07}_{-0.06}$&$16.46^{+0.10}_{-0.08}$\\
$off$&$1.0000012^{+0.0000087}_{-0.0000085}$&$1.0000055^{+0.0000125}_{-0.0000126}$&$0.0001488\pm0.0000053$&$1.0000067^{+0.0000128}_{-0.0000129}$\\
$\sigma$&$0.000149\pm0.000005$&$0.000212^{+0.000008}_{-0.000007}$&$0.000149\pm0.000005$&$0.000213\pm0.000007$\\
log~$H$&$-$&$-$&$-1.80^{+1.11}_{-0.76}$&$-1.64^{+1.18}_{-0.92}$\\
log~$B_r$&$-$&$-$&$-3.99^{+1.73}_{-1.29}$&$-3.78^{+1.52}_{-1.51}$\\
ln~$Z$&$3593.7$&$3439.8$&$3597.8$&$3454.3$\\
\toprule

\end{tabular}
\end{table}
\renewcommand{\arraystretch}{1}

\section{TACHELES model correlations} \label{TACHELES_corner}
\begin{figure}[t]
\centering
\includegraphics[width=0.9\textwidth]{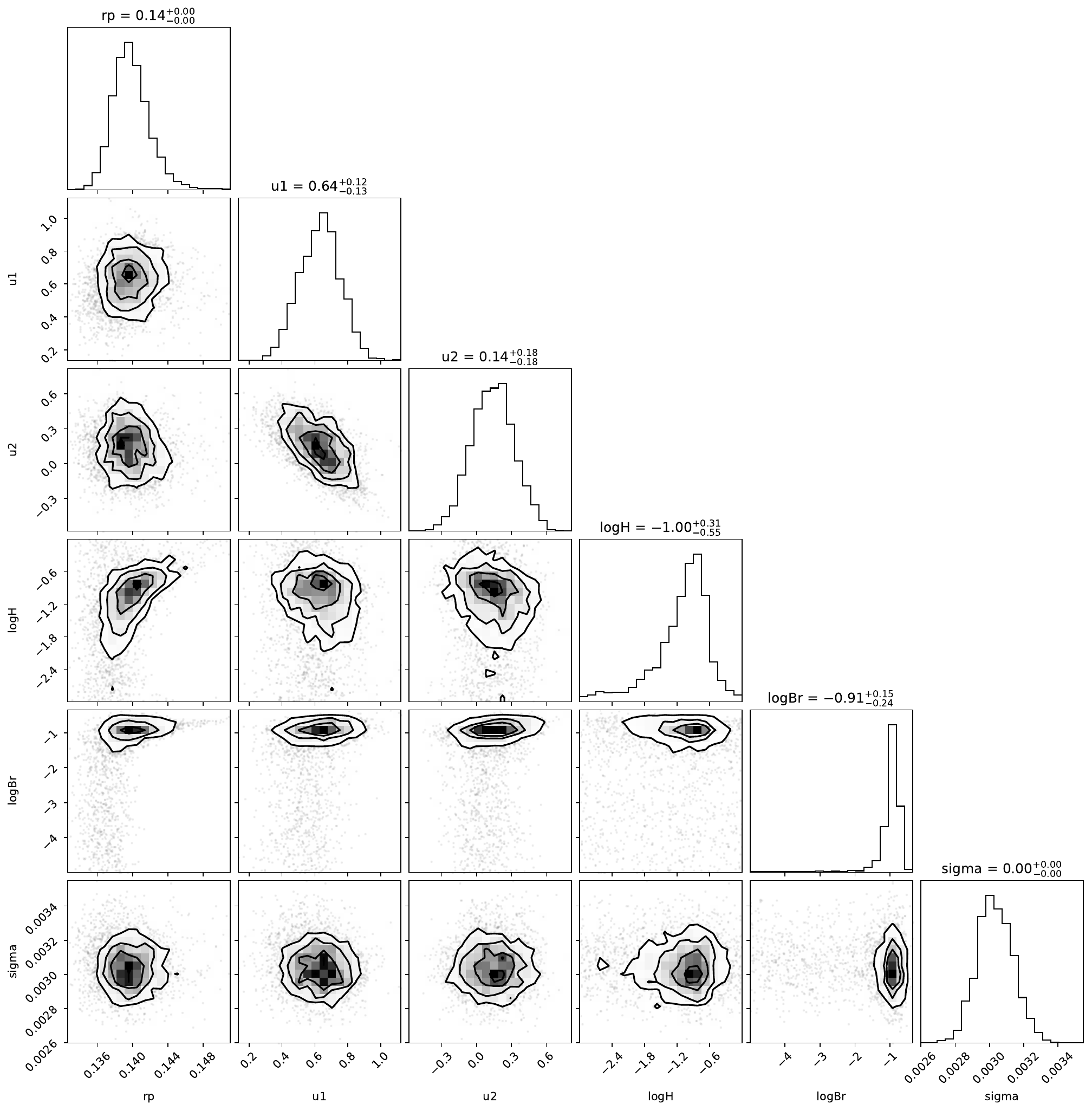}
\caption{Example corner plot of a light curve fit for a wavelength bin where H and $B_r$ are well-constrained ($\lambda=0.7516~\mu$m).}\label{fig:corner}
\end{figure}

Fig.~\ref{fig:corner} shows an example corner plot of the {\sc ULTRANEST} posterior distributions of a wavelength bin where $\log H$ and $\log B_r$ are well constrained. Fig.\ref{fig:deltaR} shows the relative radii derived by the TACHELES and photospheric models, as a function of the TACHELES brightness ratio. TACHELES's statistical advantage $\Delta ln Z$ clearly correlated with larger $B_r$ and the relative radii.

\section{Data and code availability}
All the JWST data used in this paper can be found in MAST: \dataset[https://doi.org/10.17909/2ah2-6p76]{https://doi.org/10.17909/2ah2-6p76}.
SDO images can be acquired at \url{https://sdo.oma.be}.
All derived data from this publication are available at \url{https://zenodo.org/records/13361162}. All the codes used above are also available:
\begin{itemize}
    \item {\sc TACHELES} is available on {\sc github}: \url{https://github.com/vperdelwitz/TACHELES}.
    \item {\sc POSEIDON} is available at \url{https://poseidon-retrievals.readthedocs.io/en/latest/content/installation.html}.
    \item {\sc CHIANTI} is available at \url{https://chianti-atomic.github.io}.
    \item {\sc Astropy} is available at \url{https://www.astropy.org}\citep{2022ApJ...935..167A,2018AJ....156..123A,2013A&A...558A..33A}. 
    \item  {\sc PyAstronomy} is available at \url{https://github.com/sczesla/PyAstronomy.git}. 
    \item {\sc batman} is available at \url{https://lkreidberg.github.io/batman/docs/html/index.html}.
\end{itemize}

\bibliography{Perdelwitz}{}
\bibliographystyle{aasjournal}

\end{document}